\documentclass[a4paper,twocolumn,11pt,preprint]{quantumarticle}
\pdfoutput=1
\usepackage[utf8]{inputenc}
\usepackage[english]{babel}
\usepackage[T1]{fontenc}
\usepackage{amsmath}
\usepackage{hyperref}
\usepackage{ulem}
\usepackage{graphicx}
\usepackage{tabularx}
\usepackage{array}
\usepackage{amsthm}
\usepackage{amsfonts}
\usepackage[numbers,sort&compress]{natbib}
\usepackage{comment}
\usepackage{slashbox}

\usepackage{tikz}
\usepackage{lipsum}

\newtheorem*{remark}{Remark}

\newcommand{\cc}{c.c.}

\newcommand{\pder}[2]{\frac{\partial\, #1}{\partial\, #2}}

\newcommand{\ket}[1]{\left|#1\right\rangle}
\newcommand{\bra}[1]{\left\langle#1\right|}

\title{Quantum Adversarial Learning in Emulation of Monte-Carlo Methods for Max-cut Approximation: QAOA is not optimal}

\author{Cem M.~Unsal}
\email{cem@terpmail.umd.edu}
\affiliation{Quantum Artificial Intelligence Laboratory, NASA Ames Research Center, Moffett Field, California 94035, USA}
\affiliation{KBR, 601 Jefferson St., Houston, TX 77002, USA}
\affiliation{Applied Mathematics \& Statistics, and Scientific Computation, University of Maryland, College Park, Maryland 20742, USA}
\orcid{0000-0002-7426-9702}
\thanks{\\This work was authored by employees of KBR, Inc. under Contract No. 80ARC020D0010 with the National Aeronautics and Space Administration. The United States Government retains and the publisher, by accepting the article for publication, acknowledges that the United States Government retains a non-exclusive, paid-up, irrevocable, worldwide license to reproduce, prepare derivative works, distribute copies to the public, and perform publicly and display publicly, or allow others to do so, for United States Government purposes. All other rights are reserved by the copyright owner.}
\author{Lucas T.~Brady}
\affiliation{Quantum Artificial Intelligence Laboratory, NASA Ames Research Center, Moffett Field, California 94035, USA}
\affiliation{KBR, 601 Jefferson St., Houston, TX 77002, USA}
\orcid{0000-0001-7696-7689}
\begin{document}
\maketitle

\begin{abstract}
One of the leading candidates for near-term quantum advantage is the class of Variational Quantum Algorithms, but these algorithms suffer from classical difficulty in optimizing the variational parameters as the number of parameters increases.
Therefore, it is important to understand the expressibility and power of various ans\"atze to produce target states and distributions.
To this end, we apply notions of emulation to Variational Quantum Annealing and the Quantum Approximate Optimization Algorithm (QAOA) to show that QAOA is outperformed by variational annealing schedules with equivalent numbers of parameters.
Our Variational Quantum Annealing schedule is based on a novel polynomial parameterization that can be optimized in a similar gradient-free way as QAOA, using the same physical ingredients.
In order to compare the performance of ans\"atze types, we have developed statistical notions of Monte-Carlo methods.
Monte-Carlo methods are computer programs that generate random variables that approximate a target number that is computationally hard to calculate exactly.
While the most well-known Monte-Carlo method is Monte-Carlo integration (e.g. Diffusion Monte-Carlo or path-integral quantum Monte-Carlo), QAOA is itself a Monte-Carlo method that finds good solutions to NP-complete problems such as Max-cut.
We apply these statistical Monte-Carlo notions to further elucidate the theoretical framework around these quantum algorithms.
\end{abstract}

\section{Introduction}
The idea of quantum computing was conceived in the early 1980s \cite{Feynman1982} as a way of simulating quantum systems using the laws of quantum mechanics.
Quantum simulation algorithms were further developed in the 1990s \cite{lloyd1996universal,abrams1997simulation,brown2010using,nielsen2002quantum,georgescu2014quantum} and onward generating a vast literature as one of the primary categories of quantum algorithms.
Today, these algorithms are capable of simulating many scientifically relevant systems including chemistry\cite{aspuru2005simulated}, local interactions\cite{low2017optimal}, high-dimensional quantum systems\cite{kim2017holographic,bauer2016hybrid,foss2021holographic}, lattice models\cite{wiese2013ultracold}, scattering\cite{gustafson2019quantum}, spin chains\cite{chen2010quantum}, Ising Hamiltonians\cite{edwards2010quantum}, energy transfer\cite{mcardle2019digital}, Rydberg Atoms\cite{weimer2011digital}, Unruh thermal radiation\cite{hu2019quantum}, Schwinger model\cite{kuhn2014quantum} and sparse Hamiltonians \cite{aharonov2003adiabatic}.

The goal of quantum simulation is to predict quantum properties of the simulated system such as ground state properties, binding energies, reaction rates and quantum dynamics.
Many algorithms for these purposes use quantum simulation as a subroutine to predict quantum states at a low number (relative to problem size) of different points in time.
Simulation achieves this by approximately taking a path in Hilbert space that corresponds to the path taken by the simulated system.
The correspondence must be efficiently computable for the simulation algorithm to be useful.
However, this does not necessarily mean that the quantum computer takes the most efficient path to the final state in terms of computational resources such as time, energy, hardware size, circuit depth, circuit size, or cost of gate-set \cite{bukov2019geometric}.

During the Noisy Intermediate-Scale Quantum (NISQ) \cite{preskill2018quantum} computing era one of the most important constraints on quantum circuits is the coherence time of the underlying devices.
Therefore, when generating quantum states using NISQ computers, it is important to take a time-efficient path to get there.
Quantum simulators follow a path analogous to the path of the original system, but it is often possible to take a faster path (e.g. geodesics in the control landscape \cite{bukov2019geometric}). 
The problem is that finding such faster algorithms requires greater knowledge not just of the control landscape but of the target state itself, stunting the development of such algorithms in the literature compared to their simulation counterparts.
Using Monte-Carlo emulation, we compare parameterized quantum Max-Cut approximation algorithms and demonstrate that our novel parameterization scheme can emulate a traditional parameterization scheme, such as the Quantum Approximate Optimization Algorithm (QAOA) \cite{farhi2014quantum}, in a shorter amount of time.

This paper is organized as follows: in section \ref{sec:Background} we explain the necessary background; in section \ref{sec:Monte-carlo} we extend the concepts of simulation and emulation to computer programs that are Monte-carlo methods; in section \ref{sec:Polynomial} we present the Quantum Annealing schedule that generalizes QAOA; in section \ref{sec:Data} we present our numerical data comparing QAOA to its generalization; and lastly we conclude with remarks and future directions.

\section{Background}
\label{sec:Background}
\subsection{Quantum Simulation}

Quantum simulation was the original impetus for the proposal of quantum computers \cite{Feynman1982}.
Simulating quantum systems on classical computers is suspected to be a hard problem, and Feynman's original proposal for quantum computers was a means of solving this problem by using quantum mechanics itself to simulate quantum systems.

At its core, simulation is a procedure whereby the behavior of one difficult to control system is mirrored and captured by a simpler, controllable system. 
The advantage here is that the simulator could be smaller than the simulated system, cutting away unnecessary degrees of freedom and that the simulator is a controlled environment.
While universal quantum computers can efficiently simulate any quantum system \cite{Deutsch1985}, there are many special purpose quantum simulators designed for a small number or even one task \cite{wu2021concise}.

Simulation is one of the cornerstones of quantum algorithms, and it is expected that many near-term quantum applications will be simulation-based, especially in quantum chemistry \cite{aspuru2005simulated}. 
The main limitation of quantum simulation is that it mirrors the evolution of the underlying system.
The simulator is constrained to mimic what the original system does.
This is great if the goal is to understand the entire process, but in our setting, where the only goal is the end state, simulation is too constrictive.

For instance, QAOA was originally proposed \cite{farhi2014quantum} based on inspiration from adiabatic quantum computing (AQC).  Both algorithms have the same goal, to minimize (or maximize) some cost function, but they are allowed the freedom to go about this goal in different ways.
Thus, QAOA and AQC while related, are not simulations of each other with QAOA exhibiting very different behavior in practice and even having greater computational power for some problem classes \cite{Farhi2016,Lloyd2018}.
Our goal here in going back from QAOA to a variational polynomial quantum annealing algorithm will also not be one of simulation but rather more generally emulation.

\subsection{Emulation}
In computing, emulation is the replication of a computer system's behavior by another one \cite{guruprasad2005integrated}.
Simulation is also a type of emulation namely, the emulation of the internal state of a computer system.
The most important type of emulation is the emulation of the input-output behavior of a system.
When the term "emulation" is used without any qualifiers, it usually refers to the emulation of input-output behavior.
For the rest of this paper, we will use this convention.
Just like in simulation, resource use, such as time and memory, is a central issue with the design of emulator algorithms.
In this paper, our primary goal is to show that we can emulate QAOA with the same number of variational parameters but shorter worst case requirements for time using a novel parameterization of the Quantum Annealing schedule.

\subsection{Machine Learning Concepts}
Our definition of Monte-Carlo emulation was inspired by two primary concepts in machine learning: Generative Models and Reinforcement Learning.
In Generative Models the goal is to approximately learn a probability distribution often with the goal of efficiently sampling from the target distribution\cite{ng2001discriminative}.
Specifically, we were inspired by Generative Adveserial Learning\cite{goodfellow2020generative}.
Generative Adveserial Learning is a two part system.
On one side, there is a generator which is the type of model described above.
On the other side, there is a discriminator which gets trained to classify if a given sample is from the real target distribution or from the generative model.
The generator is then trained to not be detected by the discriminator.
By training the discriminator and the generator together, it is possible to achieve a very strong discriminator and a very strong generator.
The second concept, reinforcement learning, is a type of machine learning where the objective is to maximize a reward function\cite{kaelbling1996reinforcement}.
In particular, multi-agent reinfocement learning helps define Monte-Carlo emulation.
In multi-agent reinforcement learning, each agent tries to maximize a local reward such as gaining points in competitive sport in a physics engine environment \cite{won2021control}.
We have used these ideas to define Monte-Carlo emulation.
In our definition of Monte-Carlo emulation, we have a two agent system: the emulator and the emulated.
The emulator generates an indiscriminant distribution from emulated,
an independent computer program.
On the other hand emulated tries to generate a distribution without using unnecessary resources.
We have used this definition to compare QAOA to our novel parameterization\footnote{If QAOA literature further develops where parameter optimization can not only be performed for expectation energy but also for fidelity with states that have energy bellow a given level, these ideas can be further applied to numerically analyze emulation. 
See Appendix \ref{app:future} for how these ideas can be further applied as the literature for variational quantum algorithm optimization improves.}.

\subsection{Max-cut problem and Ising Hamiltonian}
On graphs, we can partition nodes into two sets with the set of edges between the two partitions known as a "Cut".
The cardinality of Cut, or the number of edges cut, is referred to as the cutsize, and the task in Max-cut is to find the partitioning of the system that maximizes the cutsize\cite{garey1974some}.

The Max-cut can be encoded into a problem of binary variables, $z_i$, where each variable takes on a value of $\pm1$, with $+1$ corresponding to nodes in one partition and $-1$ corresponding to nodes in the other partition.  This problem is then formulated as
\begin{equation}
    \label{eq:maxcut}
    C = \sum_{\langle i, j \rangle} z_i z_j,
\end{equation}
where $\langle i, j \rangle$ denotes nodes $i$ and $j$ that share an edge in the graph.  For a given partitioning of the nodes $z_i$ into $\pm1$ bins, this cost function $C$ will change, with each edge crossing between partitions contributing $-1$ to $C$ and each edge within a partition contributing $+1$.  Therefore, a partitioning that corresponds to the Max-cut will give the lowest possible value of $C$.  Max-cut for general graphs is known to be an NP-hard problem\cite{garey1974some}.

This computational problem is equivalent to finding the ground state of the Ising model from theoretical physics \cite{ising1924beitrag}.
In an Ising model, the variables $z_i$ in Eq.~(\ref{eq:maxcut}) represent (magnetic or quantum) spins that can either be up or down\cite{lenz1920beitrag}.  $C$ in this setting corresponds to the energy of the physical system with the lowest energy state corresponding to the partitioning of the graph that gives the maximum cut.  To look at a quantum Ising model, we can take Eq.~(\ref{eq:maxcut}) and promote the variables $z_i$ to Pauli-$z$ operators acting on the $i$th qubit.  In our following analysis of quantum annealing and QAOA, we will take our problem Hamiltonian to be just this:
\begin{equation}
    \hat{C} = \sum_{\langle i, j \rangle} J_{ij}\sigma_z^{(i)} \sigma_z^{(j)},
\end{equation}
and our mixer Hamiltonian will be a transverse field on the qubits:
\begin{equation}
    \hat{B} = -\sum_{i} \sigma_x^{(i)}.
\end{equation}
In Max-cut problems, the connectivity matrix, $J_{ij}$ would just be one on edges in the graph and zero otherwise.  The Ising model is more general and corresponds to what is called weighted Max-cut where the $J_{ij}$ matrix can take on any real values.

\subsection{Monte-Carlo Methods for Optimization Problems}
\label{sec:background:carlo}
For some problems, it is very hard, if not impossible to find an exact solution.
However, it might be much easier to find a good approximate solution for the same problem \cite{metropolis1953equation}.
Monte-Carlo Methods are algorithmic techniques for generating a random variable around the exact solution.
For many practical applications, good approximate solutions can be substitutes for the exact solution and can be obtained by sampling these random variables.
There are many classical algorithms in the literature, that are used to study quantum mechanics and other areas of physics, which are Monte-Carlo methods.
Monte-Carlo methods can also be used to approximate good solutions to optimization problems.
Some optimization problems are very hard to solve exactly, such as Max-cut which is NP-hard.
According to the Exponential Time Hypothesis, the time requirements for Max-cut make it infeasible to find the exact solution in large instances \cite{lokshtanov2013lower}.
However, there are both classical and quantum algorithms to find good solutions that give a large cut of the graph \cite{goemans1994879,mazyavkina2021reinforcement,Kadowaki_1998,farhi2014quantum}.
QAOA is one such method as it samples from good solutions of optimization problems.
It has been substantially studied in its application to the Max-cut problem \cite{guerreschi2019qaoa}.
Following the example of the literature on QAOA, we will mostly keep our discussion around application performance to Max-cut however most of our arguments apply to other optimization problems as well.

For Max-cut problems, Monte-Carlo methods generate a distribution on the support of partitions with corresponding cut sizes.
This means that the Monte-Carlo method also has a distribution on the support of cut sizes.
More generally, Monte-Carlo methods for optimization problems sample from states with some energy distribution.
The Boltzmann distribution, which describes the relationship between temperature, energy, and the probability of thermodynamic states, was the original inspiration for simulated annealing, which in turn inspired Quantum Annealing \cite{metropolis1953equation}.
Boltzmann distributions have long been used as a benchmark for Monte-Carlo methods; however, it is not always the best distributions to characterize the energy distribution of a given Monte-Carlo method.
In this paper, we define a novel method for comparing the performance of Monte-Carlo methods which looks at distributions globally rather than using the temperature from the Boltzmann fit.

\begin{remark}
    Monte-Carlo methods should not be confused with Monte-Carlo algorithms for decision problems which form a class of algorithms where the probability of the correct answer is bounded below by 2/3 for all instances of the problem.
    The set of problems solvable in polynomial time by classical and quantum computers by Monte-Carlo algorithms are called BPP and BQP respectively \cite{nielsen2002quantum}.
\end{remark}

\subsection{Variational Quantum Algorithms}

Variational Quantum Algorithms have emerged as one of the most popular classes of near-term quantum algorithms, finding use cases in optimization \cite{farhi2014quantum}, chemistry \cite{peruzzo2018variational}, imaginary time evolution \cite{McArdle_2019}, quantum machine learning \cite{Biamonte_2017}, and many other applications.  This class of algorithms relies on parameterized quantum circuits and a variational classical outer loop that optimizes the quantum circuit parameters based off measurements of quantum observables.  Two of the big draws of these variational algorithms are that a) they are slightly robust to noise since the variational loop can account for this, and b) they allow for potential quantum advantage through their variational nature rather than meticulously crafted algorithmic design.

While these variational algorithms can be parameterized using any available quantum circuit, variational ans\"atze that do not incorporate enough information about the problem, often run into so-called barren plateaus where there is insufficient direction to the optimization process \cite{mcclean2018barren}.  A way to avoid this is by using as much information about the problem instance as possible in the design.  One key example of this is the Quantum Approximate Optimization Algorithm (QAOA) which employs a bang-bang control that switches between applying a simple driver Hamiltonian, $\hat{B}$, such as a transverse field on qubits, and the problem Hamiltonian, $\hat{C}$, which encodes the energy landscape of the target problem \cite{farhi2014quantum}.  By starting at the ground state of $\hat{B}$, we can apply the bang-bang control, trying to find a state that minimizes its energy with respect to $\hat{C}$, treating the lengths of the bangs as variational parameters.

QAOA was inspired by Quantum Annealing \cite{Kadowaki1998,farhi2000quantum}, which has the same goal and uses the same Hamiltonians and initial state.  While QAOA is a variational bang-bang algorithm, Quantum Annealing relies on a continuous ramp from the driver to the problem Hamiltonian.  For long enough runtimes, the quantum adiabatic theorem \cite{Jansen_2007} ensures that such an annealing procedure will transform a system, initially in the ground state of $\hat{B}$, into the ground state of $\hat{C}$ with high probability, minimizing the energy.  While quantum annealing was originally proposed to be adiabatic, many non-adiabatic or diabatic variants and usages of Quantum Annealing have been proposed \cite{Crosson_2021} with numeric success in some specific problems but without general analytic framework.

Both these algorithms fall into a class of analog quantum algorithms which can be described by the Hamiltonian evolution
\begin{equation}
    \label{eq:control_ham}
    \hat{H}(t) = u(t) \hat{B} + (1-u(t))\hat{C},
\end{equation}
where $u(t)\in[0,1]$ is the generalized annealing schedule that takes on a bang-bang form in QAOA and a decreasing ramp form in annealing.  It would of course be possible to generalize this further by replacing $(1-u(t))\to v(t)$ and having independent controls.  Whether independent controls are implementable is dependent on the quantum system at hand, and these two models have equivalent computational power under polynomial-time reductions.  Restricting to a single control field is often easier from control and theoretical point of view, and for this paper, we will only consider such a model.

Some work has gone into the connection between QAOA and Quantum Annealing \cite{farhi2014quantum,Mbeng2019,Zhou_2020,Brady2020,Brady2021}, and it has been shown that for constrained annealing time, optimizing for energy expectation and searching on the space of all annealing schedule functions, the optimal solution to Eq.~(\ref{eq:control_ham}) takes on a bang-anneal-bang form that then approaches a smooth, adiabatic like annealing ramp in the long runtime limit.  Furthermore, it has been observed that this optimal schedule oscillates at a frequency which empirically matches the optimal bang lengths in QAOA \cite{Brady2021}. 
This correspondence to an underlying optimal curve provides a link and justification for the practice of QAOA bootstrapping.

It has been noted in several studies \cite{Zhou_2020,Pagano_2020} that QAOA parameters asymptotically form smooth curves that are independent of the number of bang-bang layers, $p$.
In other words, for a $p=10$ layer QAOA procedure, the $5$th layer's parameters will look similar to the $10$th layer's parameters from a $p=20$ QAOA procedure on the same problem.  This indicates \cite{Brady2021} that the QAOA optimization is tapping into some underlying procedure, such as the above cited optimal bang-anneal-bang procedure, and this behavior provides a useful way to bootstrap up from low $p$ QAOA to higher $p$, improving the performance.

In this work, we again utilize this connection between QAOA and the optimal procedure by creating an alternative parameterization that is able to capture more information from the optimal procedure.  We still parameterize $u(t)$, using $2p$ parameters, but instead of having those parameters be bang-bang lengths, these parameters will instead be coefficients of a higher order polynomial.  We discuss the details of this parameterization and its theoretical underpinnings in Sec.~\ref{sec:Polynomial}.

For now, we conclude this background section by discussing how our polynomial parameterization fits into the context of existing variational Quantum Annealing procedures.  The concept of customizing a quantum annealing schedule to a problem dates back to Roland \& Cerf's optimized adiabatic schedule for unstructured search \cite{Roland2002} and its subsequent generalization into the Quantum Adiabatic Brachistochrone \cite{Rezakhani2009}.

Most attempts at variational optimization of an annealing schedule have worked within the context of an adiabatic path, meaning that most attempts at variational schedules still have a ramp from an initial to a final schedule.  For instance, recently VanQver and its derivatives \cite{Matsuura2020, Matsuura2021,Susa2021} have applied variational techniques on a third catalyst Hamiltonian that is only present in the middle of the anneal.  There has been some work on variationally implementing counterdiabatic schedules \cite{Sels2017,Petiziol2018,Passarelli2020,Wurtz2022}, but these still seek to create an adiabatic path.

The work we present below is to our knowledge novel in that it attempts a variational annealing schedule with no reference to adiabaticity at all, except to provide a theoretical guarantee of its correctness in the long-time limit.  Our schedule will in that sense satisfy the conditions more of diabatic quantum annealing and will seek to emulate rather than simulate an adiabatic path.

It should also be noted that variational quantum annealing and our variational polynomial schedule do have one key deficiency compared to QAOA, namely the necessary experimental control.  QAOA just requires gates applying a single Hamiltonian for some length of time; whereas, more general variational annealing requires finer time-dependent control over what mixture of Hamiltonians is applied.  This finer control requires more experimental control and might be difficult for some near-term systems to implement.

\section{Emulation of Monte-carlo Methods}
\label{sec:Monte-carlo}
\begin{figure}
    \centering
    \includegraphics[width=.5\textwidth]{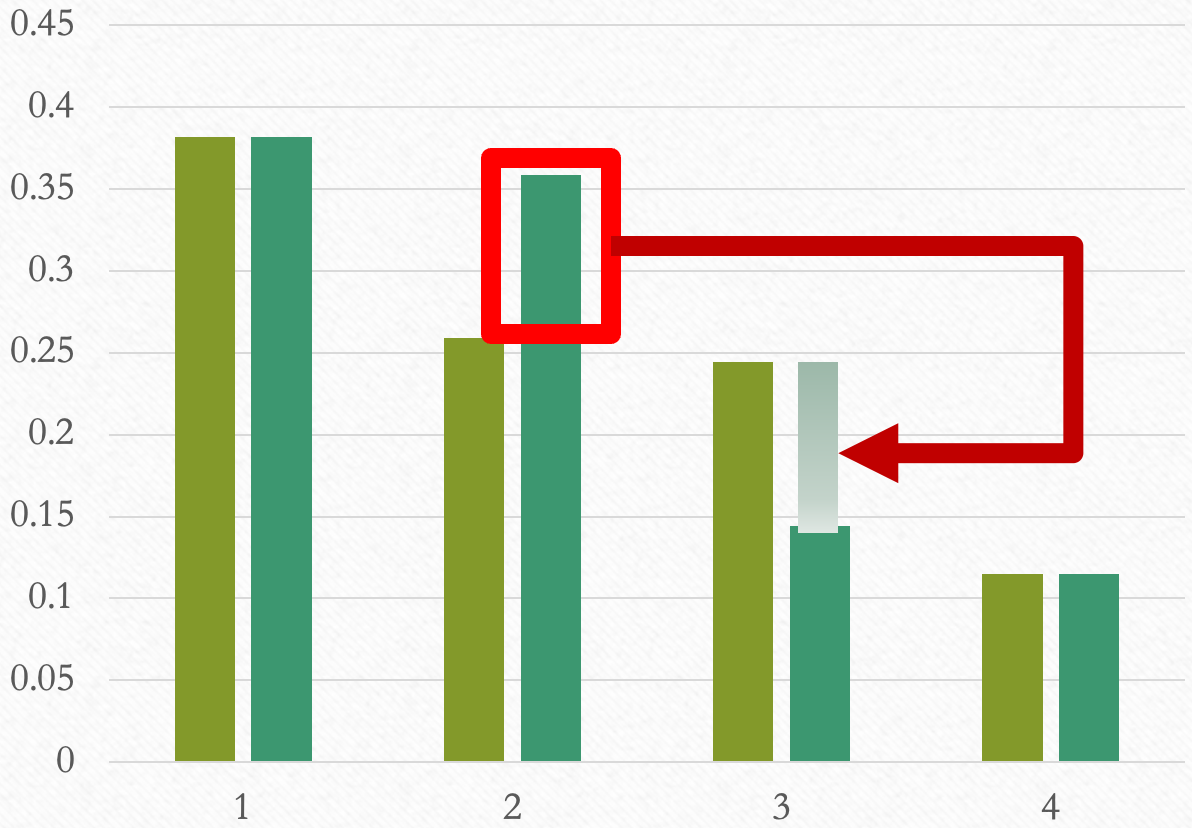}
    \caption{An example application of substitution property on a probability mass function.}
    \label{fig:sub}
\end{figure}

In this section, we extend the definition of computational emulation to computer programs that are Monte-Carlo methods.
As described in section \ref{sec:background:carlo} the energy of the solution outputs sampled from these programs is a random variable.
To replicate the behavior of the emulated, the emulator must find states with the same energy at the same rate.
Two observations help to define this.
\begin{itemize}
    \item \textbf{Interchangibility}: In optimization problems the states that have the same energy are equivalent to each other by definition.
    Therefore emulator must match the probability distribution function (pdf) of the emulated unless the following property is satisfied:
    \item \textbf{Substitution}: We assume that states that are better solutions are substitutes for states that are worse solutions however, the other way around is not true (see Figure \ref{fig:sub}).
    The probability of better solutions can be used as a substitute for the probability of worse solutions to match the pdf of the emulated distribution.
\end{itemize}

For energy minimization problems, this means the emulator must majorize the Cumulative Distribution Function (CDF) of the probabilities of getting energy eigenstates from the emulated.
Thus, for any statistic (such as median, expectation, or quantile) the performance of the emulator is bounded below by the performance of the emulated.

Monte-Carlo methods are invoked for optimization problems when the computational resources that are available is insufficient to compute the global optimum.
Therefore, they are usually adaptable to different levels of resource availability \cite{dunn2022exploring}.
When more resources are allocated, Monte-Carlo methods are usually able to produce better distributions.
These computational resources are any of the assets that needs to be provided to a computer program to run to completion such as time, memory, energy, communication network, queries etc.
By using any combination of these, we can define a cost functional.
Let
\begin{equation}
\label{eq:cost}
    c_b^g[F]
\end{equation}
be the cost of Monte-Carlo method $b$ to generate a distribution that majorizes $F$ for problem instance $g$.
The minimum cost distribution generated by $b$ does not have to match $F$ only majorize it.
Let 
$$G_b^g[F]:=c_b^g[G_b^g[F]]=c_b^g[F]$$
be such minimum cost distributions.
As a convention, let a distribution that is possible to generate but takes infinite resources, cost $\aleph_0$.
Let distributions that cannot be generated by method $b$ for problem instance $g$ cost $\aleph_1$.
For these ungenerateble distributions, method $b$ does not have $G_b^g[F]$.
We shall denote these by a function that is not a CDF
$$c_b^g[F]=\aleph_1 \iff \forall x \in \mathbb{R}, G_b^g[F](x)=2.$$

Furthermore, these definitions allows us to compare Monte-Carlo methods with each other.
In many cases, two Monte-Carlo methods could be better at different aspects in the distributions that they generate.
For example, one could have a lower expectation while the other one has lower median in its distribution.
However, given enough resources, one Monte-Carlo method might be able to emulate the other one.
To quantify the relative resource requirements of method $a$ emulating method $b$, we define "emulation factor":
$$J_{a,b}^g[F]:=\frac{c_a^g[G_b^g[F]]}{c_b^g[F]}=\frac{c_a^g[G_b^g[F]]}{c_b^g[G_b^g[F]]}.$$
For parameterized Monte-carlo methods with $p$ parameters, we define "parameterized emulation factor":
$$\hat{J}_{a,b,p}^g[F]:=\frac{c_{(a,p)}^g[G_{(b,p)}^g[F]]}{c_{(b,p)}^g[G_{(b,p)}^g[F]]}.$$
In Section \ref{sec:Data}, we will use this definition to compare bang-bang parameterization and our novel polynomial parameterization of Quantum Annelaing.
We will explore the extrema of this functional with respect to $F$, $p$ and $g$.
In the next section, we present our novel polynomial parameterization which is a generalization of QAOA that demonstrates that QAOA is not optimal by using parameterized emulation factor.

\subsection{Post-Selection on Multiple Runs}
\label{sec:amp}
One way we can improve the energy CDF of our samples is to replace each sample with the result of a binning process across $m>1$ samples picking the lowest energy in the bin to be our new sample, a process we will refer to as post-selection.
This way, the original CDF of the energy distribution of a single sample, $F(x)$, is transformed to 
$$1-(1-F(x))^m.$$
We do not include this post-selection method to improve the CDF in the scope of this study.
We choose to exclude this because, the number of queries is a computational resource such as annealing time, number of qubits, or the number of parameters.
So, one of our objectives is to avoid post-selecting on multiple samples as this would increase the number of queries to the Monte-Carlo method.
Furthermore, the number of samples needed for post-selection may be unbounded.
If there exists an $x$ value where the emulated CDF is 1 and the emulator is not 1, post-selection requires an infinite amount of samples.
This can easily be the case for large problem sizes where there is a lot of room for differing behavior between different algorithmic approaches.
Therefore, we exclude\footnote{In practice it might be possible overcome the problem to the level of machine-epsilon or other sources of noise so it could be important to study it, see section \ref{sec:future}.} post-selection from this study.

\section{Polynomial Parameterization}
\label{sec:Polynomial}
We parameterize our quantum annealing schedule by clipping polynomials.
The Clip function refers to the function which projects its input to a window \cite{virtanen2020scipy}:
\begin{equation}
    K_{a,b}(x)=min(a,max(x,b)).
\end{equation}

By scaling our Hamiltonians to the full parameter range of the experimental setup, we can ensure that the control function $u(t)\in[0,1]$.  To reflect this with the Clip function, we use $a=0$ and $b=1$.
We use a clipped linear combination of the first $2p$ monomials:
\begin{equation}
    u(t) = K_{0,1}\left(\sum_{j=0}^{2p-1}c_jx^j\right).
\end{equation}
Our use of $2p$ is so that our number of control parameters matches up with QAOA.

We optimize for the linear combination coefficients $c_j$s thus defining a variational schedule.
The $c_j$s are allowed to take any real values and, thus, do not have direct optimization bounds.

This parameterization is a generalization of QAOA because it can approximate any QAOA schedule to arbitrary accuracy by using the same number of parameters.
For a QAOA schedule with pulses changing at times $t_1,\dots,t_{2p-1}$ the corresponding $c_j$s are Lagrange interpolation of points
\begin{equation}
    (0,\lim_{y\rightarrow\infty}-y),\forall j\in \{1,2,\dots,2p-1\},(t_j,0).
\end{equation}

QAOA with $p$ layers is numerically observed to have a QAOA curve with polynomial of degree $p-1$ as defined in Refs. \cite{Zhou_2020,Pagano_2020}.
Each layer of QAOA is defined by two variables so, QAOA of depth $p$ has $2p$ variables.
This means QAOA uses $2p$ variables to traverse an underlying curve which only requires $p$ parameters to describe with polynomial parameterization.
Therefore, compared to QAOA's underlying curve, the clipped polynomial has twice the degree for the same number of variables.

\section{Comparison of Time Resource Requirements to QAOA}
\label{sec:Data}
In this section, we explore the ability of bang-bang and clipped polynomial annealing schedules to emulate one another for the same number of parameters.
We are interested in annealing time requirements for a given distribution which we will define as our cost function, (\ref{eq:cost}).
This means that we will ignore the cost of parameter optimization in this section and assume that we have an oracle which gives optimal parameters to generate $G^g_{(b,p)}$.
We have investigated the extrema of parameterized emulation factor of bang-bang and polynomial parameterizations on each other and quantified their relative capabilities.
For large problem size $|g|=n$, the summary of our results is bellow:

\begin{center}
\begin{tabular}{ | m{1.5cm} | m{2.6cm}| m{2.6cm} | } 
  \hline
   $\hat{J}_{a,b,p}^g[F]$ & $a=$Bang-Bang $b=$ Polynomial&  $a=$ Polynomial $b=$ Bang-Bang\\ 
  \hline
  \vspace{-20pt}
  $$\max_{|g|=n,F,p}$$ \vspace{-20pt}& $\aleph_1$ (section \ref{section:inemulable})& 1 (section \ref{section:lagrange}) \\ 
  \hline
  \vspace{-15pt} $$\min_{|g|=n,F,p}$$
  \vspace{-20pt}& 1 (section \ref{section:bboptimal}) &$\log(n)^{-\Omega(1)}$ (section \ref{section:simulations})\\
  \hline
\end{tabular}
\end{center}
\subsection{Emulation Limits of Bang-Bang Schedules}
\label{section:inemulable}
As formulated, both QAOA and our poly schedules are limited only by the number of parameters involved, not the length of time the procedures take.  Under this formulation, it is clear that the polynomial schedules are capable of tasks that QAOA is not.

For instance, consider a $p=1$ QAOA schedule with $2p=2$ parameters.  QAOA is known to be strictly limited in its performance by the number of parameters it has \cite{farhi2014quantum,Brady2021}, meaning that for fixed number of parameters, there is a hard limit on how close QAOA can get to the target metric.  On the other hand, even with $2p=2$ parameters, the polynomial schedule is capable of creating a linear adiabatic ramp and using the quantum adiabatic theorem to asymptotically approach the target state in the long time limit \cite{farhi2014quantum}.  Therefore, with time unbounded, the polynomial schedules can always at least perform and adiabatic path and can often perform more efficient diabatic paths as opposed to the fixed depth behavior of QAOA.

This means that QAOA can never emulate a polynomial schedule in an optimized, time unbounded setting.  This is a feature of unbounded time, and for much of our numeric study we will alleviate this issue by focusing on the minimum time for emulation, usually determined by the natural time that QAOA's parameter optimization settles to.

\subsection{Bang-Bang is a sub-type of Clipped Polynomial}
\label{section:lagrange}
As seen in section \ref{sec:Polynomial}, we can use Lagrange interpolation to approximate the Bang-Bang schedule to arbitrary accuracy.
Therefore, every Bang-Bang schedule can be asymptotically approached by a clipped polynomial schedule.
This means that the time required for a clipped polynomial to generate a certain distribution is upper bounded by the time required by the shortest time Bang-Bang schedule with the same number of parameters regardless of the problem instance.

\subsection{Bang-Bang Schedule can be the shortest schedule}
\label{section:bboptimal}

There are a few settings where QAOA-like bang-bang schedules seem to be optimal.  The most obvious example would be a single qubit setting or settings where collections of qubits act like independent qubits or effectively single spins \cite{Muthukrishnan2015,Bapat2018}.  For a single qubit, the optimal path can be shown to just be determined by standard Euler rotation angles, but this only works because of the homomorphism between $SU(2)$ and $SO(3)$.

A more interesting setting can be derived from optimal control theory which says that the optimal annealing schedule must start and end with bangs \cite{Brady2020}.  These bangs have finite size, and their size increases the shorter the amount of time the procedure is given.  Therefore, there exists runtime limit below which these two bangs dominate, resulting in a schedule that is equivalent to a depth $p=1$ QAOA schedule.  This runtime is very short, and for any sizeable system, this runtime will probably result in little success in achieving the target metric.  Still this limit does exist and can be observed numerically.  Furthermore, while the optimal annealing protocol typically does not take on an exclusively bang-bang form \cite{Brady2020}, there is nothing theoretically preventing this.
For example, in connected graph of 2 nodes, bang-bang is the optimal schedule.
Similarly, a larger graph made up of $n/2$ disconnected components of 2 nodes also has bang-bang as the optimal schedule, and it is possible that other less trivial examples exist as well.

The last setting where Bang-Bang schedules take the optimal amount of time is  asymptotically as $p\rightarrow\infty$.
This is because, with unbounded $p$, Trotter-Suzuki decomposition has no time overhead for $u(t)\in [0,1]$.
This can be seen by the standard Trotter formula which for a single time step gives
$$e^{-i\Delta t (u(t) \hat{B} +(1-u(t))\hat{C})}$$
\begin{equation}
     = 
    e^{-i\Delta t u(t) \hat{B}} e^{-i\Delta t (1-u(t))\hat{C}} + \mathcal{O}(\Delta t ^2).
\end{equation}
The operators on both sides of this equation will take time $\Delta t$, up to the corrections which will vanish in the limit $\Delta t\to0$ which can be achieved when $p\to\infty$.  Note that just because QAOA can become optimal via this small time step Trotterization does not at all mean that this is how QAOA behaves in practice.

\subsection{Computational results for time requirements for Clipped Polynomial to Majorize QAOA}
\label{section:simulations}
Unlike the previous sections, the theory in this area is harder to develop.
Hence, we rely on numerics to develop insights into how polynomial parameterization can be used to emulate a Bang-bang parameterization.
In this section, we present how low parameterized emulation factor can get for Polynomial when emulating Bang-Bang schedules.
In other words, we find the time overhead in using Bang-Bang parameterization compared to Polynomial parameterization even when the distribution of Bang-Bang parameterization is well-suited for the application.

As the number of qubits grow, the number of unique optimization problems increase very fast.
For example, there are polyexponentially many unlabeled graphs (see OEIS A000088) in the number of nodes which means that there are that many unique unweighted Max-Cut problems.
This increases variety of problems thus the probability of finding a problem that is well suited for Polynomial parameterization but not well suited for Bang-Bang parameterization.
The separation between these two parameterizations might be similar to the separation between two gate sets that satisfy Solovay–Kitaev universality.
Therefore, we would like to further conjecture that the emulation factor decreases as a poly-logarithmic function of the number of qubits
$$\frac{1}{\log(n)^{\Omega(1)}}.$$

We have computational results that support this conjecture.
Bang-bang parameter optimization is known to be NP-hard for the ideal case \cite{Bittel_2021}.
However, QAOA parameter optimization literature has seen improvements in recent years to approximate the ideal schedule very well.
Without a priori knowledge, bootstrapping with Nelder-mead is widely accepted to generate schedules that give very close expectation to what is achievable with that number of parameters.
Since this method produces a locally minimum expectation, it is reasonable to assume that it also generates a minimal time schedule for its distribution or a close approximation of it.
So we majorized these QAOA schedules with poly and observed the following time seperations:
\begin{center}
    
\begin{tabular}{|c|c|c|c|c|}
  \hline
     \backslashbox{$p$}{$n$}  & 1 & 2 & 3 & 4 \\
       \hline
       1& 1 & 1 & .98 & .963 \\
     \hline
     2 & 1 & 1 & 1 & .947 \\
     \hline
     3 & 1 & 1 & 1 & 1 \\
     \hline
\end{tabular}
\end{center}

\label{sec:Opt}

See Appendix \ref{app:data} for full data.
We would like to remind the reader that these separations are likely lower bounds on parameterized emulation factor of respective instances as literature for optimizing parameters for QAOA is well-developed but literature for optimizing polynomial parameters is not explored.
We discuss how these results can be improved in the Section \ref{sec:future}.
Next, we present the performance of different gradient-free methods for polynomial parameterization.

\section{Parameter Optimization Method}

We compared gradient-free optimizers from skquant \cite{lavrijsen2019skquant} and scipy \cite{virtanen2020scipy}.
Averaging over 10 independent instances of Erdos-Reyni graphs\footnote{Erdos-Reyni AKA, $G(n,p)$ is a random graph with $n$ nodes with probability $p$ for each edge \cite{erdos1960evolution}.} for each combination of parameters, we observed the approximation ratios in Figure \ref{fig:poly_g_f}  (see appendix \ref{app:data} for full parameter information).
We can observe that Powell, BFGS, pybobyqa, and CG consistently performed better than TNC, imfill, Nelder-Mead, and snobfit.
Among those Powell was the best performer and our results suggest that it is a robust method for optimizing the Polynomial schedule regardless of the number of qubits, polynomial degree, or annealing time.

\begin{figure}[h]
    \centering
    \includegraphics[width=.5\textwidth]{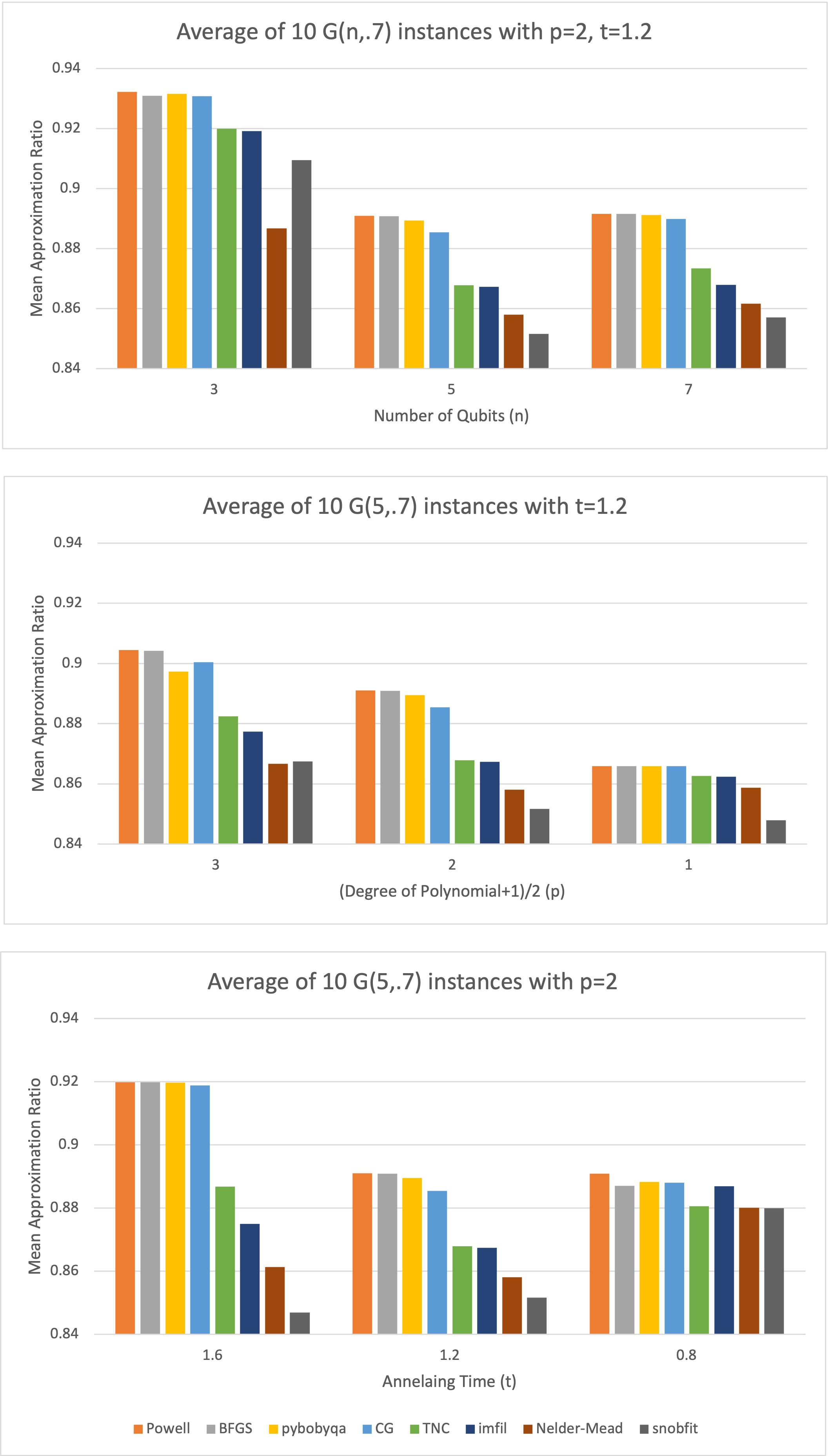}
    \caption{Comparison of gradient-free optimizers for polynomial schedule}
    \label{fig:poly_g_f}
\end{figure}

\begin{figure}[h]
    \centering
    \includegraphics[width=.5\textwidth]{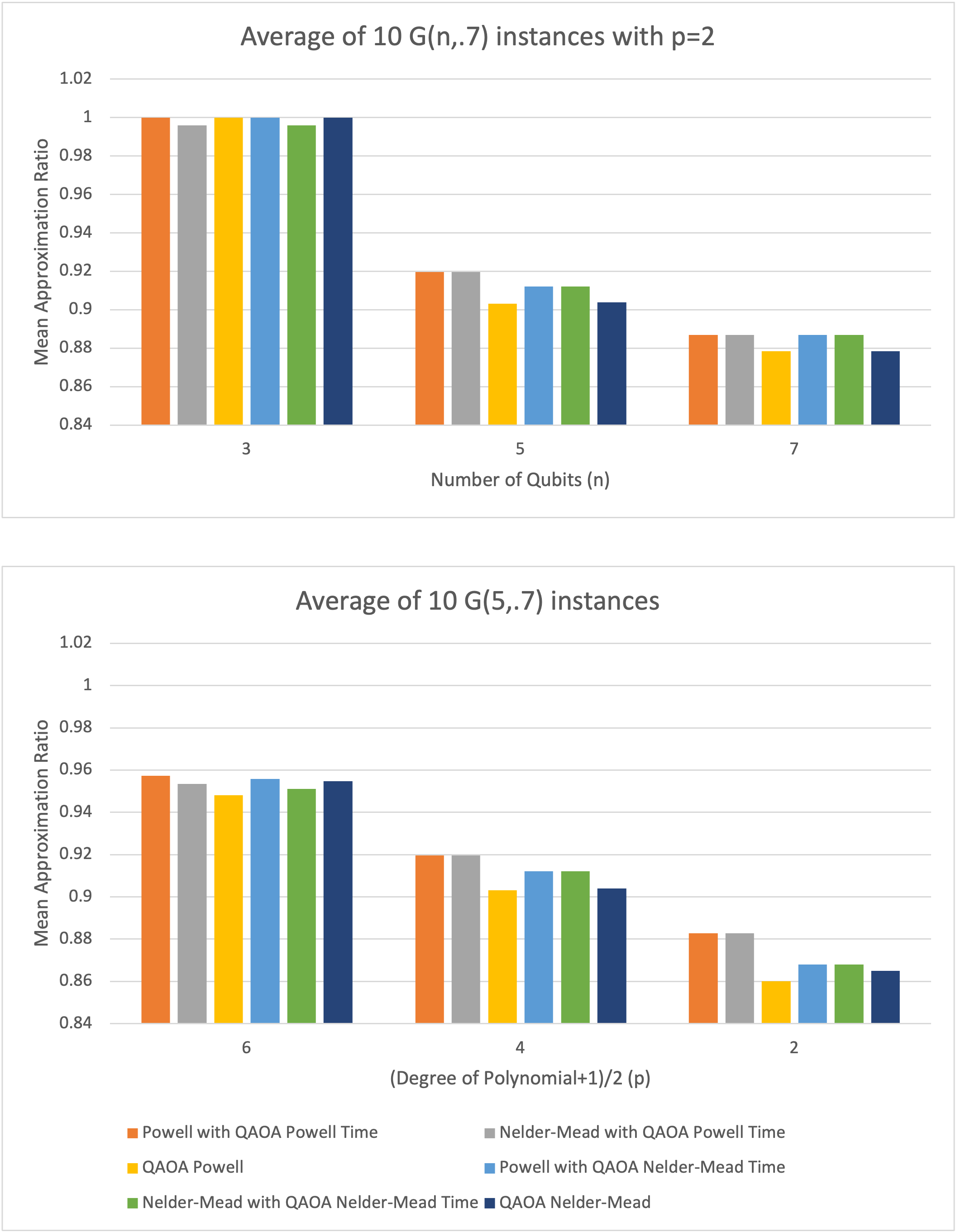}
    \caption{Comparison of gradient-free optimizers for polynomial and QAOA}
    \label{fig:qaoa_g_f}
\end{figure}

The common wisdom suggests Nelder-mead is a very good optimizer for QAOA with bootstrapping.
Unlike polynomial parameterization, annealing time in Bang-Bang schedules time is not usually treated as a computational resource.
Rather, it is determined as a consequence of optimizing the parameters.
Therefore, while optimizing for parameters if two different methods converge to different local minimum, the total time might be different.
We have first optimized QAOA optimized using Nelder-Mead and Powell.
In order to establish a fair comparison, we have taken the total QAOA time generated by both these methods and optimized Polynomial using Nelder-Mead and Powell for each of these times.
Based on the averages we collected, in Figure \ref{fig:qaoa_g_f}, we can see that Powell applied to Polynomial performs comparably to or better than QAOA regardless of the number of qubits, number of parameters, or annealing time resulting from QAOA optimization method.

\begin{figure}[h]
    \centering
    \includegraphics[width=.5\textwidth]{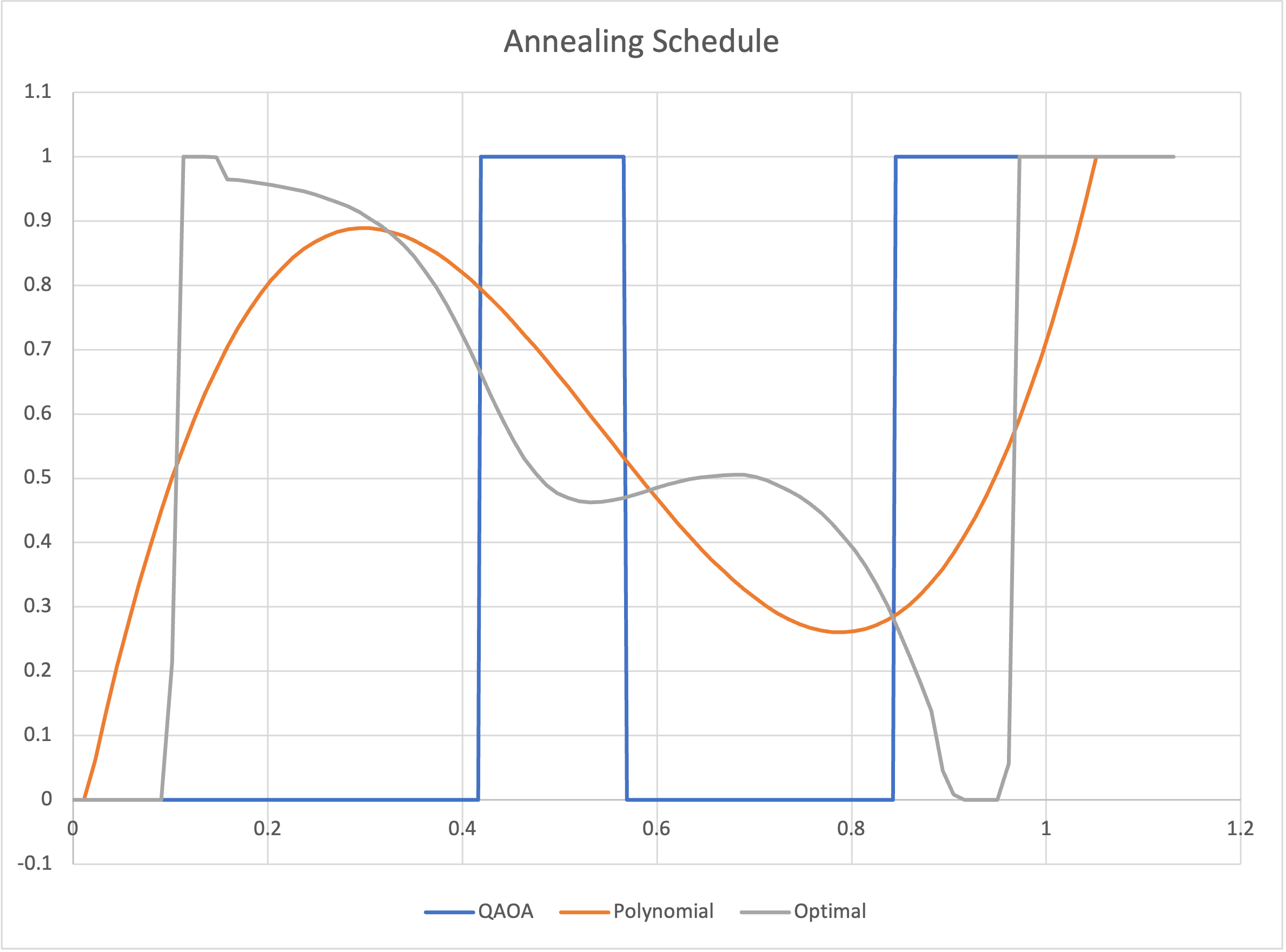}
    \includegraphics[width=.5\textwidth]{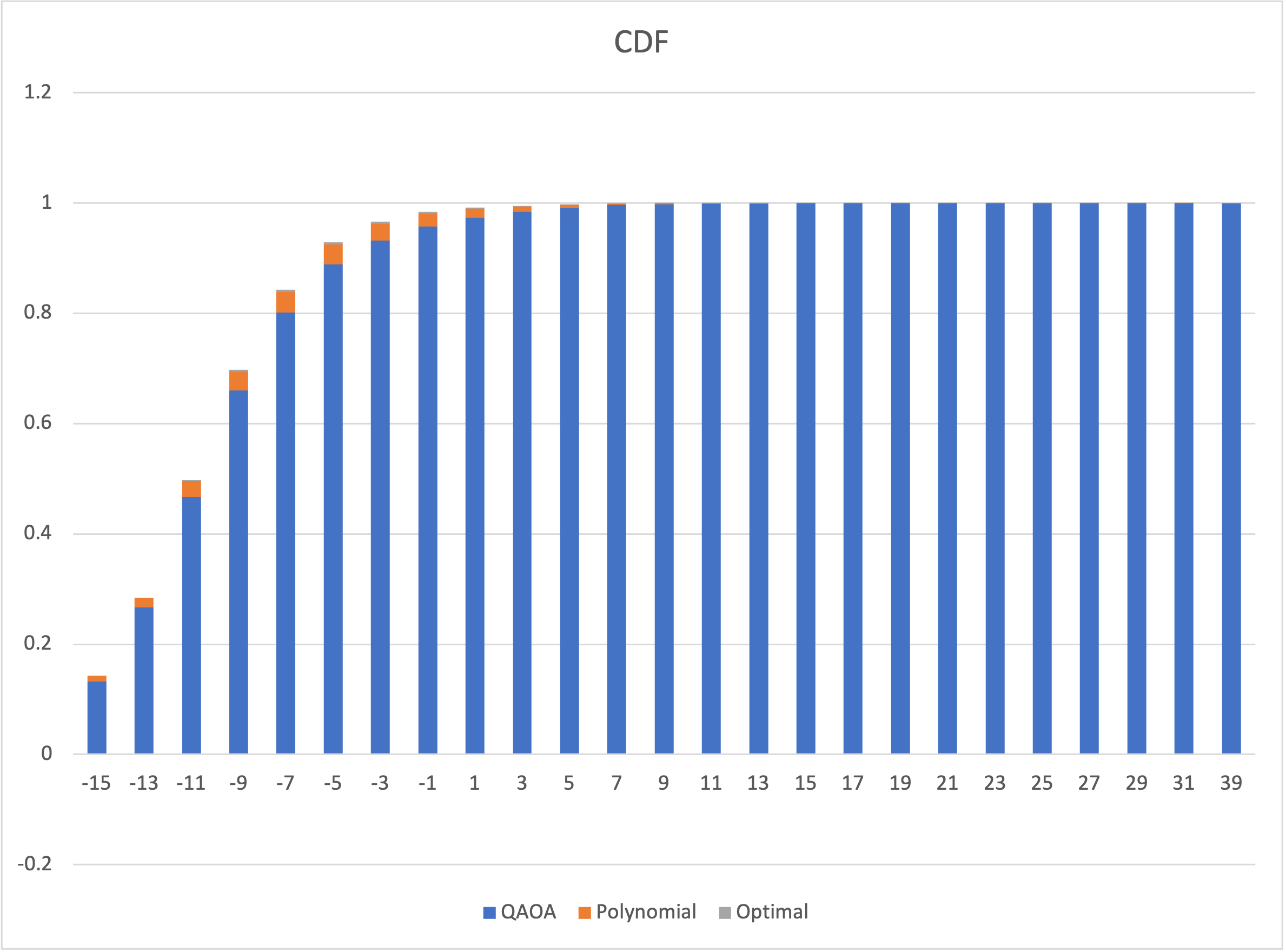}
    \caption{Resulting schedules and distributions for optimized QAOA and polynomial with 4 parameters in relation to optimal schedule.}
    \label{fig:qaoavspolyvsopt}
\end{figure}

In order to give the reader some intuition for the properties of QAOA versus the Polynomial schedules, Figure \ref{fig:qaoavspolyvsopt} shows how a Polynomial schedule optimized with Powell compares to QAOA optimized with Nelder-Mead.
We would like to note the visible difference of CDFs between QAOA and Polynomial schedules while the lack of visible difference between Polynomial and optimal schedules.
This shows that the polynomial schedule has discretized the domain of all annealing schedules effectively even with a low number of parameters.
For full data please see Appendix \ref{app:data}.

\section{Conclusion}
\label{sec:Conclusion}
To sum up, the polynomial schedule can do everything Bang-bang can do sometimes in a shorter time and never requires a longer time.
Bang-bang cannot generate some distributions that the Polynomial parametereization generates.
Polynomial can be optimized via Powell and often is a comparable or better performer in terms of approximation ratio.

All in all, the polynomial is better than the Bang-bang schedules in terms of distributions it can generate and in terms of ease of optimization.
The shortcoming of the polynomial is that the underlying computational model assumption of the Bang-bang schedule is a subset of the polynomial schedule.
Namely, the type of control required on the annealing schedule to be able to experimentally implement polynomial is at least as hard as the type of control required for QAOA.
Even for two different control models which share important properties, such as Solovay–Kitaev Universal gate sets, the experimental implementation might be vastly different and challenging in distinct ways.
Thus, we cannot assume that experimental challenges for polynomial schedule would not cause significant overhead compared to a gate-based system.

This brings us to many important implications and directions for future work.
\label{sec:future}
Firstly, this study gives motivation for experimental groups to explore ways of implementing arbitrary schedules on annealers.
Secondly, an important question about the initialization of parameters remains for polynomial.
Thirdly, Monte-Carlo emulation can and should be used in other areas of machine-learning literature to measure the relative resource requirements of different models and different types of cost functions (see Section \ref{sec:amp}).
Expanding on that, Monte-Carlo emulation gives a way for comparing the number of parameters for a given machine learning model.
Lastly, the separations that we found in the section \ref{section:simulations} were just examples.
It would be valuable to see how large these separations can be. (see appendix \ref{app:future} for suggestions to those who are interested in how this can be done).

\acknowledgements
L.T.B. and C.U. are KBR employees working under the Prime Contract No. 80ARC020D0010 with the NASA Ames Research Center.  They are grateful for support from the DARPA RQMLS program under IAA 8839, Annex 128. The United States Government retains, and by accepting the article for publication, the publisher acknowledges that the United States Government retains, a nonexclusive, paid-up, irrevocable, worldwide license to publish or reproduce the published form of this work, or allow others to do so, for United States Government purposes.

\bibliographystyle{plainnat}
\bibliography{references.bib}

\appendix
\section{Detailed Data}
\label{app:data}
For Section \ref{sec:Data}, we only considered connected graphs of certain size as disconnected graphs can be divided into smaller problems efficiently.
We have enumerated all non-isomorphic connected graphs from 0 to 4 nodes and used $2,4,6$ parameters for each case.
We picked QAOA as the Bang-Bang schedule since the literature for QAOA parameter optimization is well developed giving confidence that the choice of parameters gives a total annealing time which is optimal for the distribution it generates in Bang-Bang parameterization.
We have optimized QAOA with Nelder-Mead and bootsrapping.
We have optimized polynomial with gradient-descent at each step for the energy level that minimizes
$$F(x)/G(x)$$
where $F$ is the CDF of QAOA and $G$ is the CDF of polynomial at the current step.
We have searched for minimum annealing time where $\forall x F(x)/G(x)\ge1$ by hand.
We did not find any complete graph where poly achieved an emulation factor on QAOA that is less than 1.
However, for all of the other cases, we were able to find a value for $p$ for which QAOA was emulated faster with polynomial schedule.
The cases we found seperations are the following:\\
\begin{figure}[h]
    \centering
    \includegraphics[width=.5\textwidth]{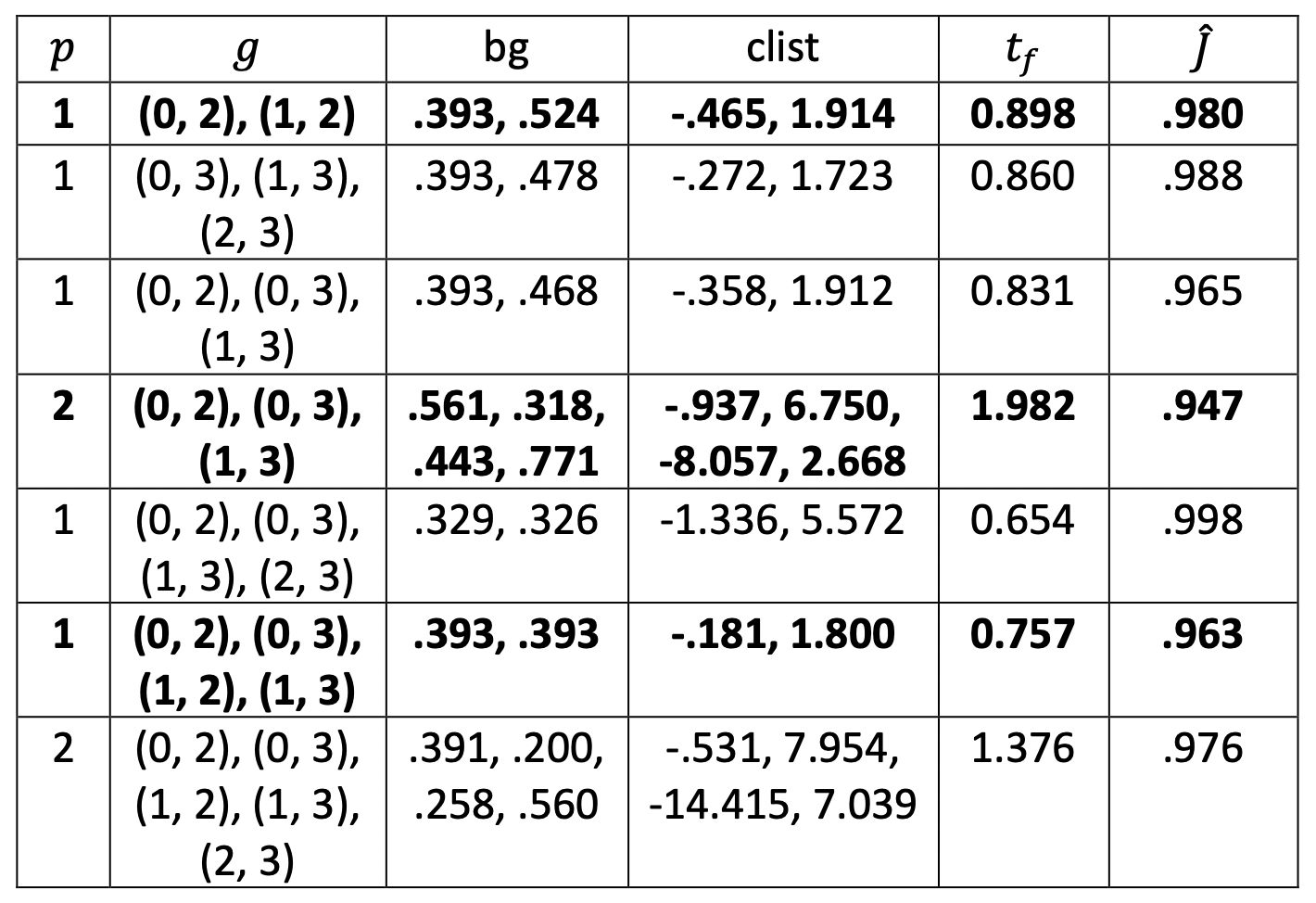}
\end{figure}
\\
where $2p$ is the number of parameters, $g$ is the graph instance, bg is the QAOA schedule as given by $\beta_1,\beta_2,\dots,\beta_p,\gamma_1,\gamma_2,\dots\gamma_p$ following the notation of Ref \cite{farhi2014quantum}, clist is the weights of the monomials in polynomial listed in increasing order by the degree, $t_f$ is the total annealing time for polynomial schedule and $\hat{J}$ is the emulation factor for polynomial on QAOA with annealing time as the cost function.
The maximum separations found for each $(n,p)$ combination is maked with bold.
We ran our simulations with 1001 points in our product formula.

For Section \ref{sec:Opt}, we have summarized our data in Figures \ref{fig:poly_g_f},\ref{fig:qaoa_g_f},\ref{fig:qaoavspolyvsopt}.
To generate the data in Figure \ref{fig:poly_g_f}, we have started with $n=5$, $p=2$ and $t=1.2$ as our baseline.
For each of the rows, we changed one parameter at a time.
For every parameter combination, we have re-sampled 10 new instances of $G(n,.7)$ with replacement and including disconnected instances.
We ran our simulations with 1001 points in our product formula.

To generate the data in Figure \ref{fig:qaoa_g_f}, we have started with $n=5$ and $p=2$ as our baseline.
For each of the rows, we changed one parameter at a time.
For every parameter combination, we have re-sampled 10 new instances of $G(n,.7)$ with replacement and including disconnected instances.
We ran our simulations with 1001 points in our product formula.

For Figure \ref{fig:qaoavspolyvsopt}, we have used the following graph: (0, 1), (0, 2), (0, 3), (0, 4), (0, 5), (0, 6), (0, 8), (0, 9), (0, 10), (0, 11), (1, 3), (1, 5), (1, 8), (1, 9), (1, 10), (1, 11), (2, 3), (2, 4), (2, 8), (2, 11), (3, 4), (3, 6), (3, 7), (3, 8), (3, 11), (4, 7), (4, 8), (4, 10), (5, 6), (5, 9), (6, 7), (6, 8), (6, 9), (6, 10), (6, 11), (7, 9), (7, 11), (8, 11), (10, 11).
We have optimized QAOA with Nelder-Mead and bootstrapping.
We have optimized polynomial parameterization with Powell.
We have optimized "Optimal" with gradient descent on 101 equally spaced points.
We ran our simulations with 101 points in our product formula.

\section{Successive Majorization}
\label{app:future}
For Section \ref{sec:Data}, we have only optimized the Bang-Bang schedule once to get QAOA and then majorized it to generate our emulation factor data.
However, these data points are likely upper bound on what how low they can be.
This is mainly due to two reasons.
Firstly, QAOA literature is well developed, giving us good confidence that the parameters are close to optimal.
Secondly, polynomial is forced to conform to Bang-Bang whereas Bang-Bang is free to optimize its expectation.
If we had a way of optimizing Bang-Bang schedule for a specific energy level and bellow as we do for polynomial, we could have had Bang-Bang and polynomial schedule iteratively majorize each other while polynomial shortens its annealing time whenever it can while keeping the majorization.
This way, not only polynomial would have conformed to Bang-Bang but also Bang-Bang would have conformed to polynomial while staying as the minimum-cost Bang-Bang schedule.
We provide our gradient descent method for polynomial schedule next.
Please note that in Eq (\ref{eq:k}), if we replace $\hat{C}$ with an indicator function for an energy level and bellow, it allows us to optimize schedule for that energy level.

\section{Optimal Control Theory}
\label{app:optimal}

In this section we will use optimal control theory to derive what the gradients are for the clipped polynomial schedule in the main text.  While we provide these results for completeness, we found in practice thatoptimization of the schedule using these gradients in general performed far worse than gradient-free optimization.

Consider a control problem of the form 
\begin{equation}
        \ket{\dot{x}(t)} = -i\hat{H}(t)\ket{x(t)},
\end{equation}
\begin{equation}
        \hat{H}(t) = u(t) \hat{B} + (1-u(t))\hat{C}.
\end{equation}

The objective function we will initially take to minimize is 
\begin{equation}
        J = \bra{x(t_f)}\hat{C}\ket{x(t_f)},
\end{equation}
but this objective function can be easily modified.

Our goal here will be to derive the conditions for optimality here given control over the function $u(t)$.  The novelty of the current approach is that we will consider the control function $u(t)$ to be a polynomial of degree $p$ so that
\begin{equation}
        \label{eq:u_poly0}
        u_0(t) = \sum_{i=0}^p c_i t^i.
\end{equation}

Officially, we would need to consider this bounded above and below so that the actual function is
\begin{equation}
        \label{eq:u_poly1}
        u(t) = \max\left(0,\min\left(1,\sum_{i=0}^p c_i t^i\right)\right).
\end{equation}
For the first portion of this derivation, we will ignore this constraint and only put it in later.

We can follow a standard optimal control procedure here \cite{Brady2020}.  Several conditions are given, but ultimately what we want is for the change in the objective function due to changes in the control function to be zero.
This condition is summarized by
\begin{equation} 
    \int_{t_0}^{t_f}\!\!dt\! \left[i\bra{x(t)}\pder{\hat{H}}{u}\ket{k(t)}+\cc\right]\delta u(t) = 0.
\end{equation}
Here 
\begin{equation} 
\ket{k(t)} = \hat{U}^\dagger(t,t_f)\hat{C}\hat{U}(t_f,0)\ket{x(0)},
\label{eq:k}
\end{equation}
where $\hat{U}(t_b,t_a)$ is the unitary time evolution operator evolving us from $t_a$ to $t_b$.  This $\ket{k(t)}$ is a Lagrange multiplier in optimal control theory, with this form being the satisfying solution to the optimal control equations for $\ket{k(t)}$.  If the target is to optimize the expectation of some goal other than $\hat{C}$, then this new target would replace $\hat{C}$ in the definition of $\ket{k(t)}$.

Naively if we did not have any constraints so that $u(t) = u_0(t)$, this could be tailored to our problem via a chain rule
\begin{equation}
        \delta u_0(t) = \sum_{i=0}^p t^i \delta c_i.
\end{equation}
By the fundamental lemma of variational calculus, we then get conditions on each $c_i$ that amount to
\begin{align} 
    \frac{\delta J}{\delta c_i}&=\int_{t_0}^{t_f}\!\!dt\! \left[i\bra{x(t)}\pder{\hat{H}}{u_0}\ket{k(t)}+\cc\right] t^i \\\nonumber
    &= \int_{t_0}^{t_f} dt\, \Phi(t)\, t^i = 0,
\end{align}
for all $i\in[0,p]$.  This $\Phi(t)$ is the gradient of the objective function with respect to the control function $u(t)$.   Therefore, the derivatives of the objective function with respect to the polynomial coefficients are just the moments of the integral of the ordinary gradient we get from optimal control.

If we want to be more precise with this, we need to consider the bounds on $u(t)$.  We can try rewriting Eq.~(\ref{eq:u_poly1}) in terms of Heaviside functions:
\begin{equation}
        \label{eq:u_poly2}
        u(t) = u_0(t) \Theta\left(u_0(t)\right)\Theta\left(1-u_0(t)\right)+\Theta(u_0(t)-1),
\end{equation}
where $u_0(t)$ is defined in Eq.~(\ref{eq:u_poly0}).

We know simply that $\frac{\delta u_0(t)}{\delta c_i} = t^i$, so we can use the chain rule to get that
\begin{align}
        \frac{\delta u}{\delta c_i} = t^i\bigg(&
                                \Theta\left(u_0(t)\right)\Theta\left(1-u_0(t)\right)\\\nonumber&
                                +
                                u_0(t) \delta\left(u_0(t)\right)\Theta\left(1-u_0(t)\right)\\\nonumber&
                                -
                                u_0(t) \Theta\left(u_0(t)\right)\delta\left(1-u_0(t)\right)+\delta(u_0-1)
                                \bigg).
\end{align}

We can simplify this a bit by recognizing that the delta functions in this expression take precedence over the Heaviside functions that will be active when the delta functions are nonzero.  Furthermore, we get another simplification from the fact that if $\delta\left(u_0(t)\right)$ is active, then $u_0(t)=0$, and similar when $u_0(t)=1$.  All of this leaves us with the simplified expression:
\begin{equation}
        \frac{\delta u}{\delta c_i} = t^i
                                \Theta\left(u_0(t)\right)\Theta\left(1-u_0(t)\right).
\end{equation}
Then the gradients for the polynomial coefficients would be

\begin{equation} 
    \frac{\delta J}{\delta c_i}= \int_{t_0}^{t_f} dt \Phi(t) t^i\Theta\left(u_0(t)\right)\Theta\left(1-u_0(t)\right) = 0,
\end{equation}

\end{document}